\documentclass[preprint2, 10pt]{aastex}

\usepackage{lscape}

\shorttitle{R Coronae Borealis stars}
\shortauthors{Hema et al.}

\begin{document}

\title{Abundance analyses of the new R\,Coronae Borealis stars: 
ASAS-RCB-8 and ASAS-RCB-10}

\author{B. P. Hema$^{1}$, Gajendra Pandey$^{1}$, Devika Kamath$^{2,3,4}$, N. Kameswara Rao$^{1,5}$, \\ David Lambert$^{5}$, and Vincent M. Woolf$^{6}$}
\affil{$^{1}$ Indian Institute of Astrophysics, Bengaluru,
    Karnataka 560034, India; hema@iiap.res.in}

\affil{$^{2}$ Department of Physics and Astronomy, Macquarie University, Sydney NSW 2109, Australia. }

\affil{$^{3}$  Australian Astronomical Observatory, PO Box 915, North Ryde, NSW 1670, Australia. }

\affil{$^{4}$ Instituut voor Sterrenkunde, KU Leuven, Celestignenlaan 200D bus 2401, B3001 Leuven, Belgium.}

\affil{$^{5}$ The W.J. McDonald Observatory, University of Texas, Austin, TX 78712-1083.}

\affil{$^{6}$ Physics Department, University of Nebraska at Omaha, Omaha, NE 68182-0266.}

\begin{abstract}

Abundance analyses of the two newly discovered R Coronae Borealis (RCB) 
stars ASAS-RCB-8 and ASAS-RCB-10 were conducted using  
high-resolution optical spectra and model atmospheres.  
Their chemical compositions place the pair
among the majority class of RCBs.   
ASAS-RCB-10 is one of the most N-poor majority RCBs with an above 
average O abundance. Relative to ASAS-RCB-10, ASAS-RCB-8 is  
H poor by 1.6 dex, O-poor by 0.7 dex but
N-rich by 0.8 dex suggesting a higher contamination by CNO-cycled material.

\end{abstract}


\section{Introduction}

R\,Coronae Borealis (RCB) stars are a rare class of 
yellow supergiants exhibiting very peculiar photometric 
and spectroscopic characteristics.
RCB stars experience  optical declines of up to several 
magnitudes at (presently) unpredictable times with 
return back to their maximum light after many days,  
months, or years. The dominant spectroscopic peculiarity is that 
their photospheres are hydrogen poor by about factors of thousands to
millions of times. RCB stars are further 
divided into  majority RCBs and minority RCBs \citep{lambert94}. 
Most notably, the  minority RCBs have  much higher
[Si/Fe] and [S/Fe] abundance ratios than the majority RCB stars.

Just over 100 RCB stars are known in the Galaxy, 
the Small Magellanic Cloud (SMC) and the Large 
Magellanic Cloud (LMC).   About 80 RCBs  have been 
identified in the Galaxy according to  
\citet{clayton96, clayton02, clayton09,
hesselbach03, miller12, tisserand13, tisserand08, zaniewski05}.
 Of the known Galactic  RCBs, 
abundance analyses are available for only about 15\%, a small sample
which may not be fully representative of the class.  
Unfortunately, some RCBs are
cool with complex spectra dominated by molecular bands, 
principally C$_2$ and CN in the optical region, and, thus, 
are a severe challenge to quantitative spectroscopy and
 many others are too faint for high-resolution spectroscopy with present telescopes.  In this paper,
we provide chemical compositions for two recently 
discovered warm and relatively bright Galactic  RCB stars.

The origin of RCB stars  is not yet without  mystery. 
The most likely  scenario  was  proposed by 
\citet{webbink84}:  a He-white dwarf merges with a 
CO-white dwarf to form the RCB with its bloated envelope 
around the C-O white dwarf. 
This scenario accounts at least qualitatively for the  
composition of the majority RCB stars including the striking discovery
 of high abundances of $^{18}$O relative to $^{16}$O
in the cooler  RCB  stars and their likely relatives, the cool H-deficient 
carbon (HdC) stars \citep{clayton05,clayton07, garc09, garc10}.
Favourable discussions of observed and predicted surface 
compositions from a merger of a He with a C-O white dwarf are provided by
\citet{pandey11,jeffery11} and \citet{menon13}. 

This paper aims to conduct an  abundance analysis for
two newly-discovered Galactic RCB stars: ASAS-RCB-8 and ASAS-RCB-10, 
found by \citet{tisserand13}. These  and other RCB stars 
were discovered from photometry using the ASAS-3 ACVS1.1\footnote{The 
ASAS-3 survey monitored the sky south of declination +28 deg 
between 2000 and 2010 to a limiting magnitude of V = 14.} 
variable star catalogue. 
 Low-resolution spectra were obtained 
to confirm that they were RCB stars.

Our abundance analysis  also considers two  other recently discovered
RCB stars V532\,Oph \citep{clayton09}  and 
V2552\,Oph {\citep{kazarovets03, kato03, hesselbach03}.  
Abundance analyses of this pair have been
previously reported: V2552\,Oph  by \citet{rao03} and 
V532\,Oph by \citet{rao14}.  New spectra
of the pair are analysed here, partly as 
a check on our analytical procedures.

\section{Observations}

High-resolution optical spectra of the program stars at their 
maximum light  were obtained with  the 1.2-m Mercator telescope 
at the Roque de Los Muchachos observatory, La Palma.
The 1.2 m Mercator telescope is equipped 
with the high-resolution fibre-fed spectrograph, 
HERMES (High-Efficiency and High-Resolution Mercator Echelle Spectrograph) 
\citep{raskin04}. The spectra were acquired at a resolving power 
($\lambda$/$d\lambda$) of 85000 over the whole wavelength domain from
3770 to 9000\AA.  Before abundance analysis, application of smoothing increased the S/N ratio and lowered the resolving power to 35000
but the intrinsically-broad stellar line profiles are barely affected.
Spectra for the program stars, ASAS-RCB-8 and ASAS-RCB-10 
were obtained along with  spectra of  V2552\,Oph 
and V532\,Oph. Spectrum of a rapidly rotating early-type star 
was also obtained for each set of observations along with 
 other calibration spectra.

McDonald observatory spectra of ASAS-RCB-8 and V2552\,Oph were 
 obtained  with the Robert G. Tull cross dispersed
echelle spectrograph of the Harlan J. Smith 2.7m reflector 
at the W. J. McDonald
observatory \citep{tull95}. The spectral resolving power 
R=$\lambda$/$d\lambda$ was 40000. The spectrum covers 3900 to
10000\AA\ with gaps beyond about 5600\AA\ where the echelle orders were 
incompletely captured on the Tektronix 2048 $\times$ 2048 CCD.  
(The McDonald spectrum of V2552\,Oph differs from that analysed   
by \citet{rao03}.)  Mercator and Tull spectra of the same object 
are well matched in strength and shape of spectral features.

\begin{table*}
\begin{center}
\caption{Log of the observations for the program stars.}
\small
\begin{tabular}{lcccc}
\tableline\tableline
Star &  Date of Observation &  V$_{obs}$ (mag)  & Obervatory & V$_{rad}$ (km $s^{-1}$)\\
\tableline
V2552\,Oph & 22 May 2007 & 11 & McDonald & 63$\pm$2 \\
V2552\,Oph & 12 May 2014 & 11 & La Palma & 60$\pm$2 \\
ASAS-RCB-8 & 11 May 2014 & 11.2 & La Palma & 134$\pm$2 \\
ASAS-RCB-8 & 23 Sep 2012 & 10.9 & McDonald & 137$\pm$1.5 \\
ASAS-RCB-10 & 12 May 2014 & 11.4 & La Palma & 54$\pm$1 \\
V532\,Oph\tablenotemark{a} & 13 May 2014 & 11.7 & La Palma & 5$\pm$1 \\
\tableline
\tablenotetext{a}{V532\,Oph's V$_{rad}$ as determined by \citet{rao14} is 
-4.6$\pm$1.5 km $s^{-1}$ and -2.7$\pm$0.9 km $s^{-1}$, respectively, for 
two different epochs (see Table 1 of \citet{rao14}).}.
\end{tabular}
\end{center}
\end{table*}

The date of observation, the AAVSO visual-validated magnitude (V)
at the time of observation and the heliocentric-corrected radial 
velocity are given in Table 1.

\section{Interstellar Extinction and Absolute Magnitude}

In this section, the interstellar reddening is estimated 
from the strengths of the diffuse interstellar bands (DIBs) 
and the distance from the  radial velocities of  the star 
and from interstellar Na D lines. Together the reddening 
and distance provide 
an estimate of a star's absolute magnitude
and this magnitude with an estimate of the stellar mass 
defines a locus in the ($T_{\rm eff},\log g$) plane which in 
conjunction with spectroscopic loci serves to constrain 
the atmospheric parameters $T_{\rm eff}$ and $\log  g$.

 DIBs  are identified in the spectra of
V532\,Oph, V2552\,Oph, ASAS-RCB-10 and ASAS-RCB-8.
Equivalent widths of seven DIBs were measured.
For estimating the reddening  E(B-V),
we have used `equivalent width -- reddening' relation given by
\citet{luna08} and \citet{herbig93}. 
The estimated average E(B-V) is 0.52, 0.76, 0.57 and 0.25 lead to  
 A$_{\rm V}$  of 1.70, 2.39, 1.78 and 0.77 
(assuming  R of 3.1) for V2552\,Oph, 
V532\,Oph, ASAS-RCB-10 and ASAS-RCB-8, respectively. The A$_{\rm V}$
values are consistent with the estimates of \citet{tisserand13}
of 1.89 and 0.62 for ASAS-RCB-10 and ASAS-RCB-8, respectively.
%based on far-infrared surveys. 
%The A$_{v}$ for V532\,Oph from our studies 
%is an excellent match with that from  \citet{rao14}.

The near maximum light magnitudes for program stars are given in Table 1. 
 V maximum values for the program stars were corrected for 
interstellar extinction. These extinction-corrected  V$_{0}$ are,
9.3$\pm$0.13 for V2552\,Oph, 9.3$\pm$0.15 for V532\,Oph and 9.62$\pm$0.15  
for ASAS-RCB-10. 
%The light maximum V magnitude for ASAS-RCB-8 has
%been estimated to be 10.9 by \citet{tisserand13}.
AAVSO V band magnitudes that are
available show a V magnitude  for ASAS-RCB-8 of
10.91$\pm$0.1. This V maximum
value may then be corrected for interstellar extinction of 0.7 mag. 
resulting in V$_{0}$ of 10.2$\pm$0.15.
The estimates of $E(B-V)$ for V532\,Oph are in excellent agreement 
with that of \citet{rao14}. But for V2552\,Oph, our determinations of 
$E(B-V)$ are different from that of \citet{rao03}.

Kinematics of the interstellar gas along the line of 
sight to ASAS-RCB-8 are  revealed by the Na\,{\sc i} D lines. 
As an aid to identifying the interstellar Na D components, 
we computed the expected radial velocity with respect to 
the local standard of rest (LSR) with distance in the direction of ASAS-RCB-8 
using the model Galactic rotation given by \citet{brand93}. 
Components with negative LSR radial velocities are not
expected to occur in this direction. The Galactic rotation 
curve shows  an almost  linear increase of V$_{\rm lsr}$ with 
distance and suggests that the radial velocity of the star 
(V$_{\rm lsr}$ $\sim$ 144 km s$^{-1}$) might occur at a distance 
of 8 kpc. Presence of ISM clouds in front of a 
star provides a minimum distance to the star of 7 kpc and    a 
lower limit to the absolute visual magnitude (M$_{V}$) of this star. 
The extinction-corrected V magnitude 10.2$\pm$0.1 obtained earlier coupled
with the distance estimate of 7 kpc provides the
minimum M$_{V}$ of -4.0$\pm$0.2. However if the star
is at a distance of 8 kpc the M$_{V}$ would -4.3. 

Similarly, the distance estimated for ASAS-RCB-10 is about 7.4 kpc for 
the star's LSR corrected radial velocity 
(V$_{\rm lsr}$) $\sim$ 66 km s$^{-1}$. 
Using the extinction-corrected V magnitude 9.62$\pm$0.15 along with the 
distance estimate, the absolute magnitude of the star would be 
$-$4.7$\pm$0.2. For V2552\,Oph, the distance estimated is about 7.7 kpc
for the star's LSR-corrected radial velocity  
(V$_{\rm lsr}$) $\sim$ 72 km s$^{-1}$. 
Using the extinction-corrected V magnitude 9.3$\pm$0.13 along with the
distance estimate, the absolute magnitude of the 
star would be $-$5.1$\pm$0.2.

The bolometric correction for stars with the 
T$_{\rm eff}$ of 6750 K is expected to be small: 
$\sim$ 0.02 according to \citet{flower96} for normal supergiants.
Our adopted M$_{bol}$ for the stars are $-4.0\pm0.2$ for 
ASAS-RCB-8 and $-4.7\pm0.2$ for ASAS-RCB-10
where the uncertainty includes an estimate arising 
from the uncertain distance determinations.  
Unfortunately, the distances
can not yet be replaced by  parallax-based estimates. 
Two of the four stars are in the presently-available 
{\it GAIA} catalogue \citep{gaia16, gaia16b} but the parallaxes 
are quite uncertain: 
$\pi$ = 0.18$\pm$0.24 mas for ASAS-RCB-10 and $\pi$ = 0.59$\pm$0.25 mas 
for V2552\,Oph.  

As is well known,  $M_{\rm bol}$ can be expressed in terms of 
effective temperature $T_{\rm eff}$, surface gravity $\log g$ 
and the stellar mass $M$ as  

$M_{bol}$ $=$ $M_{bol,\odot}$ $+$ 2.5 log ($g$/$g_{\odot}$) $-$ 10 log ($T_{\rm eff}$/$T_{\odot}$) $-$ 2.5 log ($M$/$M_{\odot}$)

%log $g$ = log $g_{\odot}$ + 4 log ($T_{\rm eff}$/$T_{\odot}$) $+$ log (M/M$_{\odot}$) $+$ 0.4($M_{bol}$ $-$ $M_{bol,\odot}$)

A mass of $M/M_{\odot}$ = 0.7$\pm$0.2 is assumed considering the
proposed scenario for the origin of RCB stars \citep{weiss87, iben96}.
This relation for a given $M_{\rm bol}$  then provides a 
locus in the ($T_{\rm eff},\log g$)  plane which (see below) 
with other loci contributes to the determination of the atmospheric 
parameters $T_{\rm eff}$ and $\log g$.

\section{Abundance Analysis}

The abundance analysis is modeled closely on the analyses 
discussed in detail by \citet{asplund00} and followed in subsequent
analyses  by  \citet{rao03} for V2552\,Oph, \citet{rao08} for 
V\,CrA and Rao et al. (2014) for V532\,Oph.  Lines selected
for measurement  were largely those chosen in these 
earlier studies with the exception that fluorine 
(F\,{\sc i}) lines were chosen following
\citet{pandey08} and the C$_2$ Swan bands were synthesized 
following \citet{hema12}.

The LTE  line analyses of the program 
stars were conducted by combining the UPPSALA 
line-blanketed H-deficient model atmospheres constructed based on 
the usual assumptions: flux constant, plane-parallel layers in
hydrostatic and local-thermodynamic equilibrium (LTE), 
with the UPPSALA equivalent width analysis program `EQWIDTH'
\citep{1997A&A...318..521A}.

In the RCB stars,
the chief source of continuum  opacity in the optical
spectra  is photoionization of neutral carbon \citep{asplund00}. 
Hence,  a line of element X  is influenced by the
abundance ratio X/C  even though He is the most common species. 
Since carbon is the dominant opacity source,
the observed C\,{\sc i} lines from levels with excitation 
potentials similar to the levels providing the continuous
opacity are insensitive to the
stellar parameters. Indeed, the strength of a given C\,{\sc i} 
line is expected to be of
same strength in    spectra of RCB stars of different stellar
parameters, while the strength of the other
elemental spectral lines will vary with the change in 
stellar parameters and the metallicity \citep{rao96}.  

These expectations are confirmed but for one troubling surprise;  
the predicted strengths of C\,{\sc i} lines are weaker than observed.
Equivalently, the carbon abundance derived using the 
H-deficient line-blanketed
model atmospheres is  a factor of 4 or about 0.6 dex
less than that adopted for the
input model atmosphere. This discrepancy is dubbed  the
`carbon problem' by \citet{asplund00}.   Further insights 
into the carbon problem's resolution are potentially provided by
observations of  [C\,{\sc i}]  lines  by \citet{pandey04} 
and the C$_2$ Swan bands \citep{hema12}.  \citet{asplund00} 
proposed an array of  resolutions with some now failing to 
satisfy constraints set by the [C\,{\sc i}] and C$_2$ observations.
Their favored  resolution supposed that  the photosphere of 
Nature's RCB stars exhibited  a shallower temperature gradient than
the photosphere emerging from a computer. In principle, this 
idea might also satisfy the [C\,{\sc i}] and C$_2$ observations.
As a temporary but probably adequate measure, Asplund et al. 
suggested that abundance ratios such as O/Fe might be little
affected by the inability to resolve the carbon problem. 
Thus in what follows our focus is on abundance ratios. The abundance
analysis  is based on model atmospheres 
constructed for a C/He ratio of 1\% by number, a value
close to the mean of 0.6\% for extreme helium stars (EHes)
(see Table 1 of \citet{jeffery11}). 
EHes are sufficiently hotter than RCB stars
that helium provides the continuous opacity and, hence, 
the C/He ratio is  measurable from the strengths of He and C
lines. EHes and RCB stars are likely on very similar evolutionary tracks. 

An appropriate model atmosphere from the Uppsala grid 
is selected using a series of indicators each providing 
a locus in the ($T_{\rm eff}, \log g)$ plane. A measure of 
$T_{\rm eff}$ with only a slight dependence on $\log g$ 
is set by demanding excitation balance using permitted lines 
of Fe\,{\sc ii}, and  permitted 
and forbidden lines of O\,{\sc i}. 
The lower excitation potential for Fe\,{\sc ii} and O\,{\sc i} lines 
range from 2.8eV to 6.2eV and 0eV to 11eV, respectively.
 Ionisation balance set using
 lines of Fe\,{\sc i} and Fe\,{\sc ii}, 
Mg\,{\sc i} and Mg\,{\sc ii}, and  Si\,{\sc i} and Si\,{\sc ii} 
provided three loci.
Finally, the locus corresponding to the M$_{\rm bol}$ estimate  
for a mass of 0.7M$_{\odot}$
provides the final locus. To conclude the selection of parameters,
 lines of C\,{\sc i}, Fe\,{\sc i}, 
N\,{\sc i}, Ca\,{\sc i}, S\,{\sc i} and Si\,{\sc i} were used 
for separate determinations of the microturbulence, a quantity 
essentially independent of effective temperature and surface gravity.

Since both V532\,Oph and V2552\,Oph have been analysed previously, 
we compare our results with these previous
estimates. Our determinations of the stellar parameters for the V2552\,Oph,
for the spectra from McDonald  and  La Palma
observatories are an excellent match and are in 
agreement with those adopted by
\citet{rao03}. From our analysis, the determined stellar parameters
for V2552\,Oph are, ($T_{\rm eff}$, log $g$, $\xi_{t}$):
(6750$\pm$250, 0.4$\pm$0.5, 7.5$\pm$1).  \citet{rao03} 
adopted parameters (6750, 0.5, 7). Abundances for C/He = 1\% models
are compared in Table 2.

For V532\,Oph, we compared the equivalent widths measured 
from the La Palma spectrum
with those measured by \citet{rao14} from  their McDonald 
spectrum. Since these were an excellent match, 
we adopted the stellar parameters derived by \citet{rao14}.
They are, ($T_{\rm eff}$ K, log $g$ (cgs units), $\xi_{t}$ km\,s$^{-1}$):
(6750$\pm$250, 0.5$\pm$0.3, 7.5$\pm$1.0).

The abundances from the previously published results and from this work
are given in Table 2 for the program stars V2552\,Oph and V532\,Oph. 
The derived abundances in this study are in good agreement with 
that of the previously published results, except for minor differences
in the abundances which are attributable to the errors in the
measured equivalent widths and differences in the lines used.
The  consistency between our abundances and 
previously published results shows that our results for
ASAS-RCB-8 and ASAS-RCB-10 will be on the scale established 
by \citet{asplund00}.

\begin{figure}
\epsscale{1.0}
\plotone{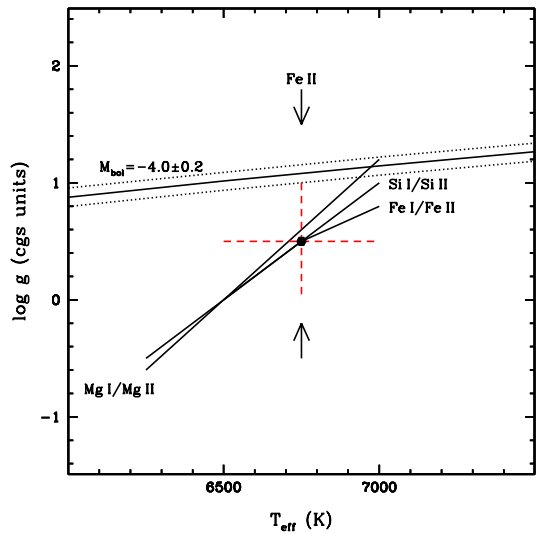}
\caption{The $\log g$ vs. $T_{\rm eff}$ plane is shown for ASAS-RCB-8.
Black up-down arrows mark  the locus of the excitation balance for the 
Fe\,{\sc ii} lines. The loci of the ionization balance for the Mg\,{\sc i} 
and Mg\,{\sc ii}, Si\,{\sc i} and Si\,{\sc ii}, and Fe\,{\sc i} and 
Fe\,{\sc ii} are shown.  The locus  for the determined 
M$_{\rm bol}$ = $-$4.0$\pm$0.2 is also shown along with the loci for
the errors on M$_{\rm bol}$ in dotted lines. 
The black dot and the red-dashed cross shows the final $T_{\rm eff}$ -- 
$\log g$ for the star.  
\label{}}
\end{figure}

\begin{figure}
\epsscale{1.0}
\plotone{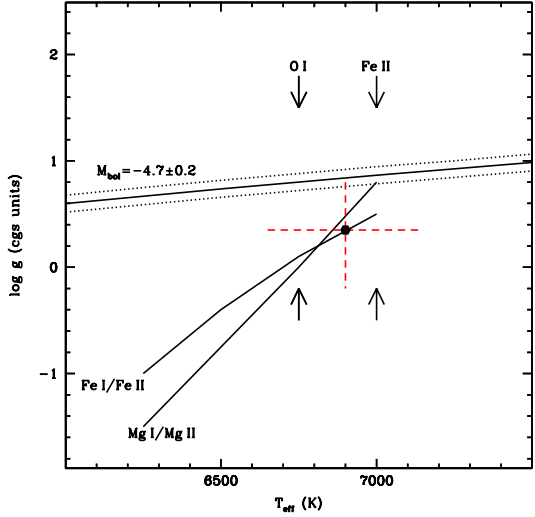}
\caption{The $\log g$ vs. $T_{\rm eff}$ plane is shown for ASAS-RCB-10.
Black up-down opposite arrows mark  the loci of the excitation balance for the
Fe\,{\sc ii} lines, and O\,{\sc i} and [O\,{\sc i}] lines. 
The loci of the ionization balance for the Mg\,{\sc i}
and Mg\,{\sc ii} and Fe\,{\sc i} and Fe\,{\sc ii} are shown. 
The locus  for the determined
M$_{\rm bol}$ = $-$4.7$\pm$0.2 is also shown along with the loci for
the errors on M$_{\rm bol}$ in dotted lines.
The black dot and the red-dashed cross shows the final $T_{\rm eff}$ --
$\log g$ for the star.  
\label{}}
\end{figure}

The stellar parameters determined for the new
RCB stars, ASAS-RCB-8 and ASAS-RCB-10 respectively are 
($T_{\rm eff}$ K, log $g$ (cgs units), $\xi_{t}$ km\,s$^{-1}$): 
(6750$\pm$250, 0.5$\pm$0.5, 7$\pm$1) and 
(6900$\pm$250, 0.35$\pm$0.5, 7$\pm$1). 
Figure 1 and 2 show the $\log g$ vs. $T_{\rm eff}$ plane 
for ASAS-RCB-8 and ASA-RCB-10,  respectively. 
Adopted  stellar parameters are indicated by a dot 
and  error bars with the  cross.  Abundances are summarized in Table 3.
The errors on elemental abundances given are the mean of 
line-to-line scatter of the abundances. The errors 
on the abundances due to  the errors on the 
stellar parameters are similar to 
those calculated by \citet{asplund00} 
(see Table 4 of \citet{asplund00}). 
The effects of $\Delta$$T_{\rm eff}$, $\Delta$$\log g$, $\Delta$$\xi_{t}$
on the abundance ratios, (X/H) and (X/Fe) is less than 0.1 dex.

\begin{figure}
\epsscale{1.0}
\plotone{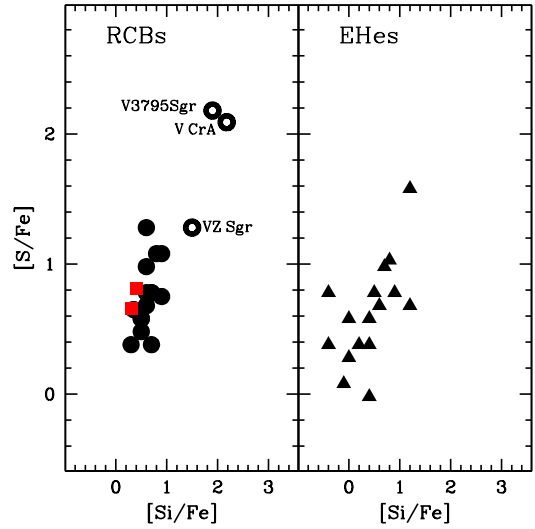}
\caption{The abundance ratios of [Si/Fe] vs [S/Fe] are given. 
The left panel shows the plot for RCB stars along with the two 
new RCBs ASAS-RCB-8 and ASAS-RCB-10
in red squares. The right panel shows the plot for EHes. 
\label{}}
\end{figure}

\section{Chemical composition}

By all principal abundance indicators, ASAS-RCB-8 and 
ASA-RCB-10 belong to the majority class and not the minority
class of RCB stars.  Figure 3 (left-hand panel) 
shows abundance ratio [Si/Fe] versus [S/Fe] for RCB stars 
analysed by \citet{asplund00} and
others.  Figure 3 (right-hand panel) shows these abundance 
ratios for EHe stars with  data taken from (\citet{jeffery11}
and the references therein).
  
ASAS-RCB-8 and ASAS-RCB-10 are positioned firmly in the majority 
clump, as are V532\,Oph and V2552\,Oph, as noted earlier by 
\citet{rao03} and \citet{rao14}.  
Certain minority RCB stars V3795 Sgr and V\,CrA are very 
cleanly separated from the majority stars.  
VZ\,Sgr also previously designated a minority RCB is
less cleanly separated from majority RCBs. The spread in 
[Si/Fe] and [S/Fe] for RCBs and EHes appears to
exceed the spreads expected from the abundance determinations.   
In previous discussions - for example, that by \citet{rao08} -- 
DY\,Cen, which has been termed a hot RCB,  appears as a 
minority RCB star. However, this appearance rested on a 
low Fe abundance \citep{jeffery93}
but  reanalysis by \citet{pandey14} including consideration 
of non-LTE effects places this hot RCB with the majority RCB stars.  
The EHe sample
does not provide a minority star but they are spread out 
more than the RCB stars  in [Si/Fe] and [S/Fe] along a 
line of a roughly constant Si/S  ratio which is
displaced about 0.4 dex from  a similar line for the RCB stars.

The majority RCBs have similar but not identical compositions.  
In Figure 4 and 5 we show histograms for several elements with the 
pair of ASAS-RCB stars identified and minority RCBs distinguished 
from the majority stars.  For the majority RCBs, the width of the 
histograms exceeds the  likely errors with the largest widths 
occurring for H and Li (not shown).
The 4 dex spread for H is obvious from the star-to-star 
differences in the appearance of the H$\alpha$ line.  
Lithium is strongly present in a few stars and clearly absent 
in most including ASAS-RCB-8 and ASAS-RCB-10. 
 For N and O, the histograms for majority RCBs spans about 1.5 dex. 
The Y abundances span
about 2 dex.   Inspection of Figure 5 shows that the Fe 
abundances are centered about the abundance 6.5 or a deficiency of 1 
dex relative to the Sun.  Given the carbon problem (see above),  
the Fe deficiency (and similar results for other elements) may arise 
from use of an  inappropriate atmospheric structure. However,
the deficiency of Fe by about 1.0 dex relative to solar is 
roughly consistent with the `kinematic' abundance and supports
the choice of C/He ratio of 1\% for RCBs (for details see \citet{rao96}).
Insistence of the same C/He ($= 1.0\%$ by number)  across the 
sample may also have contributed to the spread in the histograms;  
for example, use of C/He $= 10\%$  increases
most abundances by 1.0 dex, i.e., would remove the Fe underabundance.  
However, the adopted 1\% C/He ratio is already almost a factor of two greater 
than the measured ratio for EHes stars, possible close relatives of the RCBs,
as noted by \citet{raonk08}.

Within the majority RCB population ASAS-RCB-10 has a 
low N  and a high O abundance for a N/O ratio of $-0.9$ dex. 
ASAS-RCB-8 has  the high N/O ratio of $+0.6$ dex. This reversal 
of  the N/O ratio may indicate that ASAS-RCB-8 is now 
contaminated more severely than ASAS-RCB-10 with CNO-cycled
H-burnt material.  Consistent with this speculation is 
the lower H abundance of ASAS-RCB-8 and its generally higher 
metal abundance (the lower C abundance leads to lower continuous 
opacity and stronger metal lines, all being equal).

\begin{figure}
\epsscale{1.0}
\plotone{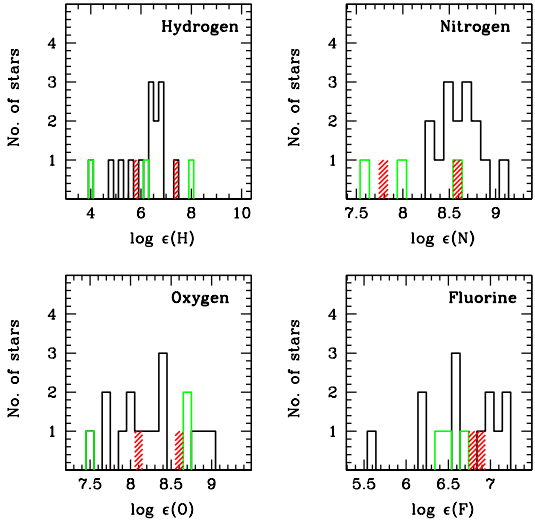}
\caption{The frequency histograms of the elemental abundances 
for H, N, O and F are shown
for the majority and minority and the two new RCB stars. 
Black line represents the majority  RCB stars and
green line the three  minority RCB stars, and 
red hatched lines represent the new  RCB stars ASAS-RCB-8 and ASAS-RCB-10.
% along with in the blue and green thick lines, respectively. 
%The new RCB stars ASAS-RCB-8 and ASAS-RCB-10 are shown in red lines.
\label{}}
\end{figure}

\begin{figure}
\epsscale{1.0}
\plotone{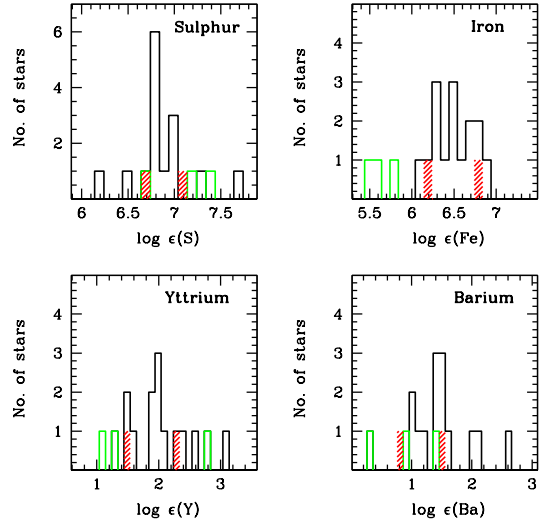}
\caption{The frequency histograms of the elemental abundances of 
S, Fe Y, and Ba are shown 
for the majority and minority and the new RCB stars.  
Black line represents the majority RCB stars, 
green line represents the three minority RCB stars, and 
red hatched lines represent the new RCB stars ASAS-RCB-8 and ASAS-RCB-10.
%The histograms shows the abundances for the majority and  
%minority RCB stars in the blue and green thick lines, respectively. 
%The new RCB stars ASAS-RCB-8 and ASAS-RCB-10 are shown in red lines.
\label{}}
\end{figure}

Spectra of all stars contain the C$_2$ Swan bands with and 
the (0,0), (0,1) and (1,0) band heads at 5165, 5636 and 4737 \AA,
respectively. Spectrum synthesis was used to fit the observed 
bands - see Hema et al. (2012) for details.  
The derived carbon abundance from C$_2$ Swan bands
for the new stars confirms the results of 
our previous study \citep{hema12},
that these derived abundances are independent of the
adopted model atmosphere's carbon abundance.
The derived carbon abundances from the (0,1) C$_2$ band
for V2552\,Oph, V532\,Oph, ASAS-RCB-8 and ASAS-RCB-10 are 
8.1$\pm$0.1, 8.2$\pm$0.1, 8.3$\pm$0.1, 8.2$\pm$0.1, 
respectively, for C/He ratio 1\%. 
The C abundance provided by the Swan bands and 
from C\,{\sc i} lines is compromised by the lack of 
understanding of the carbon problem (see above). 
In principle, the (1,0) band with the $^{12}$C$^{13}$C
head displaced to the red of the blue-degraded $^{12}$C$_2$ 
could provide an estimate of the $^{12}$C/$^{13}$C ratio. 
Unfortunately, a Fe\,{\sc i} line is blended with the 
$^{12}$C$^{13}$C head. 
 However,  the $^{12}$C$^{13}$C band head 
is almost reproduced by this Fe\,{\sc i} line, indicating that 
there is very little or no $^{13}$C present in these stars.   
The high lower bound to the   $^{12}$C/$^{13}$C ratio
is as expected for the merger product.
%and no useful information can be provided about the isotopic ratio.

\clearpage
%\begin{center}
\thispagestyle{empty}
%\begin{landscape}
%\hoffset -0.5 in
%\voffset -1.5 in
\begin{deluxetable}{lccccccccccc}
%\tabletypesize{\scriptsize}
\tablecolumns{18}
\tablewidth{-40pc}
\tablecaption{The derived and the previously published elemental abundances 
for the RCB stars V2552\,Oph and V532\,Oph.}
\tablehead{
 \colhead{} & \multicolumn{4}{c}{log$\epsilon$(E) for V2552\,Oph} & \colhead{} &  \multicolumn{4}{c}{log$\epsilon$(E) for V532\,Oph} \\
\cline{2-5} \cline{7-10}
\colhead{Elements} & \colhead{Now\tablenotemark{a}} & \colhead{n\tablenotemark{a,d}}  & \colhead{2003\tablenotemark{b}} & \colhead{n\tablenotemark{b,d}}   & \colhead{} & \colhead{Now\tablenotemark{a}} & \colhead{n\tablenotemark{a,d}} &  \colhead{2014\tablenotemark{c}} & \colhead{n\tablenotemark{c,d}} }
\startdata
H\,{\sc i} & 6.67 & 1 & 6.66 & 1 && 6.57 & 1 & 6.31 &  1 \\
Li\,{\sc i} & \nodata & \nodata & $<$0.96 & 1 && \nodata & \nodata & $<$0.97 & 1 \\
C\,{\sc i} & 9.0$\pm$0.26 & 31 & 9.05$\pm$0.24 & 17  && 8.82$\pm$0.3 & 14 & 8.91$\pm$0.34 & 14\\
%C$_{2}$\tablenotemark{d} & 8.1$\pm$0.1 & 3 & \nodata & \nodata && 8.2$\pm$0.1 & 3 & \nodata & \nodata\\
N\,{\sc i} & 8.80$\pm$0.3 & 17 & 8.96$\pm$0.28 & 8 && 8.57$\pm$0.23 & 10 & 8.57$\pm$0.23 & 10\\
O\,{\sc i} & 8.0$\pm$0.20 & 10 & 7.96$\pm$0.21 & 6 && 8.03$\pm$0.25 & 5 & 7.90$\pm$0.34 & 6\\
Ne\,{\sc i}& \nodata  & \nodata & \nodata & \nodata && \nodata & \nodata & 8.62$\pm$0.21 & 3 \\
F\,{\sc i} & 6.7$\pm$0.04 & 4 & \nodata & \nodata && 6.55$\pm$0.08 & 5 & \nodata& \nodata\\
Na\,{\sc i}& 5.95$\pm$0.28 & 3 & 6.14$\pm$0.12 & 3 && 6.16$\pm$0.13 & 4 & 6.22$\pm$0.04 & 4\\
Mg\,{\sc i}& 6.8$\pm$0.03 & 2 & 6.77$\pm$0.09 & 2 && 6.8$\pm$0.14 & 3 & 6.84$\pm$0.10 & 3\\
Mg\,{\sc ii}& 6.9 & 1 & \nodata & \nodata && 6.8 & 1 & 6.89 & 1\\
Al\,{\sc i} & 5.9$\pm$0.18 & 5 & 5.78$\pm$0.11 & 2 && 5.6$\pm$0.23 & 4 & 5.81$\pm$0.15 & 4\\
Al\,{\sc ii}& \nodata & \nodata & \nodata & \nodata && \nodata & \nodata & 5.81$\pm$0.15  & 4\\
Si\,{\sc i} & 6.74$\pm$0.3 & 11 & 6.97$\pm$0.22 & 5 && 6.95$\pm$0.23 & 7 & 6.93$\pm$0.21 & 7\\
Si\,{\sc ii}& 7.0$\pm$0.18 & 2 & 7.67$\pm$0.28 & 2 && 6.82 & 1 & 6.82 & 1\\
S\,{\sc i} & 7.0$\pm$0.24 & 8 & 6.77$\pm$0.09 & 7 && 6.65$\pm$0.27 & 7 & 6.76$\pm$0.23 & 7\\
K\,{\sc i} & \nodata & \nodata & 4.48 & 1 && 4.73 & 1 & 4.83 & 1\\
Ca\,{\sc i}& 5.2$\pm$0.3 & 13 & 5.34$\pm$0.27 & 4 && 5.13$\pm$0.23 & 6 & 5.19$\pm$0.21 & 6\\
Sc\,{\sc ii} & 2.8$\pm$0.23 & 3 & 2.91$\pm$0.23 & 3 && 2.57$\pm$0.3 & 3 & 2.80$\pm$0.27 & 3\\
Ti\,{\sc ii} & 4.15$\pm$0.20 & 3 & 4.11$\pm$0.37 & 2 &&  4.18$\pm$0.12 & 2 & 4.21$\pm$0.26 & 2\\
Fe\,{\sc i} & 6.50$\pm$0.25 & 42 & 6.37$\pm$0.27 & 21 &&  6.44$\pm$0.12 & 29 & 6.47$\pm$0.21 & 31\\
Fe\,{\sc ii} & 6.56$\pm$0.2 & 10 & 6.42$\pm$0.09 & 10 && 6.44$\pm$0.12 & 10 & 6.51$\pm$0.13 & 10\\
Ni\,{\sc i} & 5.63$\pm$0.3 & 8 & 5.55$\pm$0.18 & 5 && 5.57$\pm$0.13 & 5 & 5.55$\pm$0.13 & 5\\
Cu\,{\sc i} & 4.36$\pm$0.02 & 2 & 3.95$\pm$0.16 & 2 && 4.33$\pm$0.08 & 2 & 4.08$\pm$0.20 & 2\\
Zn\,{\sc i} & \nodata & \nodata & 4.31 & 1 && \nodata & \nodata & 4.41 & 1\\
Sr\,{\sc ii}& \nodata  & \nodata & 4.28 & 1 && \nodata & \nodata & \nodata & \nodata \\
%Y\,{\sc i} & \nodata & \nodata & \nodata & \nodata &&  \nodata & \nodata & \nodata &\nodata \\
Y\,{\sc ii}& 2.4$\pm$0.15 & 5 & 2.29$\pm$0.09 & 3 && 1.96$\pm$0.06 & 5 & 2.08$\pm$0.09 & 5\\
%Zr\,{\sc i}& \nodata  & \nodata & \nodata &\nodata &&  \nodata & \nodata & \nodata & \nodata \\
Zr\,{\sc ii} & 2.5$\pm$0.13 & 2 & 2.25$\pm0.19$ & 3 && 2.07$\pm$0.08 & 3 & 2.11$\pm$0.15 & 3\\
Ba\,{\sc ii} & 0.8$\pm$0.1 & 2 & 1.03$\pm$0.09 & 3 && \nodata & \nodata & 1.45$\pm$0.05 & 5\\
La\,{\sc ii} & \nodata & \nodata & 0.63$\pm$0.43 & 2 &&  \nodata & \nodata & 0.67$\pm$0.30 & 2\\
\enddata
\tablecomments
{$^{a}$ Using the spectra from La Palma Observatory. \\
$^{b}$ from \citet{rao03}\\
$^{c}$ from \citet{rao14} \\
$^{d}$ n is the number of lines used in the analysis.}
\end{deluxetable}

\clearpage
%\begin{center}
\thispagestyle{empty}
%\begin{landscape}
%\hoffset -0.5 in
%\voffset -1.5 in
\begin{deluxetable}{lcccccccc}
%\tabletypesize{\normal}
\tablecolumns{9}
\tablewidth{-40pc}
\tablecaption{The derived elemental abundances for the new RCB stars 
ASAS-RCB-8 and ASAS-RCB-10.}
\tablehead{
 \colhead{} & 
\multicolumn{2}{c}{ASAS-RCB-8\tablenotemark{a}} & \colhead{} &
\multicolumn{2}{c}{ASAS-RCB-8\tablenotemark{b}} & \colhead{} &
\multicolumn{2}{c}{ASAS-RCB-10\tablenotemark{a}} \\
\cline{2-3} \cline{5-6} \cline{8-9}
\colhead{Elements} & \colhead{log$\epsilon$(E)}   & \colhead{n\tablenotemark{c}}  && \colhead{log$\epsilon$(E)} & \colhead{n\tablenotemark{c}}  && \colhead{log$\epsilon$(E)} & \colhead{n\tablenotemark{c}}}
\startdata
H\,{\sc i} & 5.83 & 1 && 5.83 & 1 && 7.4 & 1 \\
Li\,{\sc i} & \nodata & \nodata && 0.96 & 1 && \nodata & \nodata  \\
C\,{\sc i} & 9.0$\pm$0.25 & 23 && 8.94$\pm$0.27 & 13 && 9.1$\pm$0.25 & 31 \\
%C$_{2}$\tablenotemark{c} & 8.3$\pm$0.1 & 3 && \nodata & \nodata && 8.2$\pm$0.1 & 3\\
N\,{\sc i} & 8.58$\pm$0.28 & 16 && 8.71$\pm$0.29 & 11 && 7.8$\pm$0.3 & 13 \\
O\,{\sc i} & 8.09$\pm$0.15 & 7 && 8.03$\pm$0.26 & 7 && 8.66$\pm$0.27 & 10 \\
Ne\,{\sc i}& \nodata & \nodata && 8.53$\pm$0.27 & 2 && \nodata & \nodata \\
F\,{\sc i} & 6.87$\pm$0.07 & 3 && \nodata & \nodata && 6.75$\pm$0.15 & 4 \\
Na\,{\sc i}& 6.35$\pm$0.04 & 2 && 6.32$\pm$0.05 & 3 && 5.92$\pm$0.07 & 3 \\
Mg\,{\sc i}& 6.79$\pm$0.02 & 2 && 6.86$\pm$0.75 & 4 && 6.83$\pm$0.22 & 3  \\
Mg\,{\sc ii}& 6.71 & 1 && 6.51 & 1 && 6.7 & 1 \\
Al\,{\sc i} & 6.1$\pm$0.2 & 4 && 5.89$\pm$0.17 & 5 && 5.6$\pm$0.26 & 4\\
Al\,{\sc ii}& \nodata & \nodata && 7.49 & 1 && \nodata & \nodata \\
Si\,{\sc i} & 7.1$\pm$0.27 & 10 && 7.5$\pm$0.2 & 13 && 6.56$\pm$0.27 & 10 \\
Si\,{\sc ii}& 7.15$\pm$0.28 & 3 && 7.22$\pm$0.4 & 3 &&  7.0$\pm$0.08 & 3 \\
S\,{\sc i} &  7.08$\pm$0.29 & 9 && 7.0$\pm$0.14 & 8 && 6.65$\pm$0.19 & 7 \\
K\,{\sc i} &  4.93$\pm$0.18 & 2 && 4.6$\pm$0.3 & 2 && 4.68 & 1 \\
Ca\,{\sc i}& 5.60$\pm$0.3 & 13 && 5.45$\pm$0.37 & 9 && 5.1$\pm$0.28 & 12 \\
Sc\,{\sc ii} & 3.03$\pm$0.3 & 4 && 2.74$\pm$0.3 & 4 && 2.55$\pm$0.24 & 3 \\
Ti\,{\sc ii} & 4.38$\pm$0.26 & 3 && 4.29$\pm$0.15 & 3 &&  3.83$\pm$0.25 & 3 \\
Fe\,{\sc i} & 6.78$\pm$0.20 & 37 && 6.57$\pm$0.4 & 39 &&  6.25$\pm$0.27 & 38 \\
Fe\,{\sc ii} & 6.80$\pm$0.13 & 15 && 6.46$\pm$0.07 & 11 && 6.2$\pm$0.29 & 12 \\
Ni\,{\sc i} &  5.94$\pm$0.17 & 7 && 5.69$\pm$0.24 & 8 && 5.33$\pm$0.30 & 3 \\
Cu\,{\sc i} & 4.43$\pm$0.23 & 2 && 4.11$\pm$0.17 & 2 && 4.24$\pm$0.25 & 2 \\
Zn\,{\sc i} & \nodata & \nodata && 4.5 & 1 && \nodata & \nodata \\
Sr\,{\sc ii}& \nodata & \nodata && 2.28 & 1  && \nodata & \nodata \\
%Y\,{\sc i} & \nodata & \nodata && \nodata & \nodata && 5.11 & 1\\
Y\,{\sc ii}& 2.33$\pm$0.3 & 7 && 2.24$\pm$0.14 & 5 && 1.5$\pm$0.12 & 4 \\
%Zr\,{\sc i}& \nodata & \nodata && \nodata & \nodata && 3.98 & 1 \\
Zr\,{\sc ii} & 2.56$\pm$0.3 & 4 && 2.27$\pm$0.14 & 3 && 2.1$\pm$0.04 & 2\\
Ba\,{\sc ii} &  1.5$\pm$0.3 & 3 && 1.1$\pm$0.2 & 3 && 0.80$\pm$0.27 & 3 \\
La\,{\sc ii} & \nodata & \nodata && 0.89$\pm$0.31 & 2 &&  \nodata & \nodata \\
\enddata
\tablecomments
{$^{a}$ Using the spectra from LaPalma Observatory. \\
$^{b}$ Using the spectra from McDonald  Observatory.\\ 
$^{c}$ n is the number of lines used in the analysis.}
\end{deluxetable}

\section{Circumstellar dust}

By chemical composition, ASAS-RCB-8 and ASAS-RCB-10 are 
representative of the majority class of RCB stars as
defined by \citet{lambert94}. Within that class there is 
an apparent spread in composition including some real 
star-to-star differences. ASAS-RCB-8 and ASAS-RCB-10 are 
examples of these differences includng H, N and O.

Production of obscuring soot clouds with severe H-deficiency 
completes the pair of defining characteristics of RCB stars.  Judged by
their soot content, i.e., their infrared excess, the two new 
RCB stars are also representative of the majority class. 
The recent investigation of Spitzer spectra by \citet{rao15} explored 
the characteristics of the  infrared excesses  of 
RCB stars  \citep{hernandez11, hernandez13}. 
The spectral energy distribution (SED) for a star was constructed
from optical and infrared photometry corrected for interstellar 
reddening.  The SED was then fitted with a blackbody to 
represent the stellar flux and additional blackbodies to
account for the circumstellar dust emission.

A SED for ASAS-RCB-8 was constructed for E(B-V) = 0.23 
with optical photometry from AAVSO and near-infrared photometry
from DENIS and 2MASS (quoted in \citet{tisserand13}). 
Mid-infrared photometry is from the satellite WISE \citep{wright10} 
in 2010.  Blackbody fits to the stellar fluxes 
and the infrared emission by dust gives a stellar blackbody 
temperature of 7200 K and dust temperatures of 400 K.
The fraction of stellar flux absorbed and reemitted
by the dust is  expressed by the covering factor R = 0.005 
in 2010. This value of R is
among the lowest values for RCB stars and probably not a surprise 
as \citet{tisserand13} remark `only one decline was observed during 
the ten-year long ASAS-3 survey. With 0.003 mag.day$^{-1}$, the
decline rate is about ten times slower than that
of classical RCB stars'. 

Similarly for ASAS-RCB-10,  blackbody fits for the stellar fluxes 
and the infrared emission by dust were made. 
Since the  observed infrared fluxes from WISE and AKARI were different, 
 separate fits were made. The derived dust temperatures are:
780 K and 730 K for WISE and AKARI observations, respectively. 
The infrared flux was lower at the time of AKARI observations
relative to the WISE observations  but the variability of the  
infrared flux is common among RCB stars.
The R values for these observations are 0.20 from WISE and 0.13 from
AKARI. The mean R value of 0.16 is taken for ASAS-RCB-10.

The amount of dust produced recently by the star (i.e., the factor R)
was checked for correlation with the photospheric elemental 
abundances by \citet{rao15}. 
The factor, R, indicating the amount of dust production shows
correlation with the hydrogen, oxygen, and also carbon
(Figures 6). R is 
 anti-correlated with Fe (Figure 6). 
The new RCB stars ASAS-RCB-8 and ASAS-RCB-10 fall on the 
trends indicated  by the previously known RCB stars and, in particular,  
anchor more firmly the low R end of the correlations.

\begin{figure}
\epsscale{1.0}
\plotone{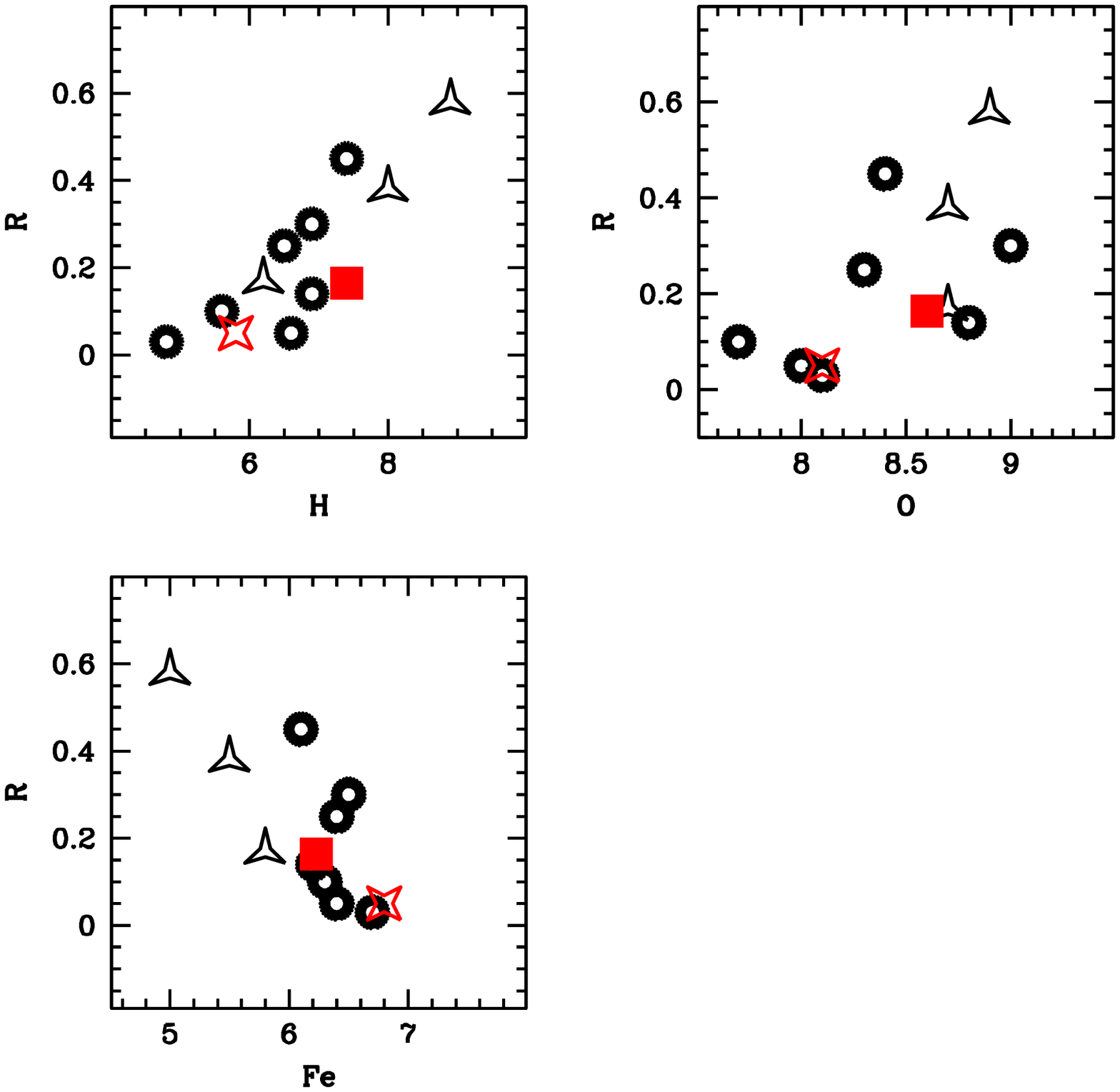}
\caption{The dust covering factor R vs. the abundances of 
H,  O,  and Fe. The majority RCB stars are shown in 
black open circles, minority RCBs  with open triangles, ASAS-RCB-8 
with open red square and ASAS-RCB-10 with filled red square. 
\label{}}
\end{figure}

\section{Concluding Remarks}

The  RCB stars ASAS-RCB-8 and ASAS-RCB-10 are new additions  
to the group of majority RCBs having the
elemental abundances similar to those of previously  identified 
majority RCB stars. In addition, these new RCB stars
support the correlations previously  reported between stellar 
composition and the strength of the emission by the circumstellar dust.

DLL acknowledges support from the Robert A. Welch Foundation of Houston, 
Texas through grant F-634. NKR would like to thank
Satya and Vimala Mandapati for their kind hospitality 
during his stay in Austin. DK acknowledges the support of 
the KU Leuven contract GOA/13/012. DK acknowledges the support 
by the Fund for Scientific Research of Flanders grant G.OB86.13.

Based on observations obtained with the HERMES spectrograph, 
which is supported by the Research Foundation - Flanders (FWO), 
Belgium, the Research Council of KU Leuven, Belgium, the Fonds 
National de la Recherche Scientifique (F.R.S.-FNRS), Belgium, 
the Royal Observatory of Belgium, the  Observatoire de Geneve, 
Switzerland and the Thringer Landessternwarte Tautenburg, Germany.

%\bibliographystyle{apj}
%\bibliography{ref}

\end{document}